\documentclass[twocolumn,showpacs,preprintnumbers,pra,runinaddress]{revtex4}
\usepackage{amssymb}
\usepackage{amsmath}
\usepackage{graphicx}
\usepackage{dcolumn}
\usepackage{bm}
\usepackage{appendix}

\newtheorem{theorem}{Theorem}

\begin{document}

\title{Apply current exponential de Finetti theorem to realistic quantum key
distribution}

\begin{abstract}
In the realistic quantum key distribution (QKD), Alice and Bob respectively
get a quantum state from an unknown channel, whose dimension may be unknown.
However, while discussing the security, sometime we need to know exact
dimension, since current exponential de Finetti theorem, crucial to the
information-theoretical security proof, is deeply related with the dimension
and can only be applied to finite dimensional case. Here we address this
problem in detail. We show that if POVM elements corresponding to Alice and
Bob's measured results can be well described in a finite dimensional
subspace with sufficiently small error, then dimensions of Alice and Bob's
states can be almost regarded as finite. Since the security is well defined
by the smooth entropy, which is continuous with the density matrix, the
small error of state actually means small change of security. Then the
security of unknown-dimensional system can be solved. Finally we prove that
for heterodyne detection continuous variable QKD and differential phase
shift QKD, the collective attack is optimal under the infinite key size case.
\end{abstract}

\author{Yi-Bo Zhao}\email{zhaoyibo@mail.ustc.edu.cn}
\affiliation{Key Lab of Quantum Information, University of Science and Technology of
China, (CAS), Hefei, Anhui 230026, China}
\author{Zheng-Fu Han}
\affiliation{Key Lab of Quantum Information, University of Science and Technology of
China, (CAS), Hefei, Anhui 230026, China}
\author{Guang-Can Guo}
\affiliation{Key Lab of Quantum Information, University of Science and Technology of
China, (CAS), Hefei, Anhui 230026, China}
\maketitle

\section{Introduction:}

Information-theoretical security proof \cite{Renner thesis} is a powerful
and general way to prove the security for quantum key distribution (QKD). In
this method, to give the amount of unconditional secret keys, we only need
to discuss upper or lower bounds of some entropies. The exponential de
Finetti theorem is crucial to this method, which support that as the key
size goes to infinite, Eve cannot get more information from the coherent
attack than from the collective attack \cite{Renner thesis}. Since in the
collective attack Eve attacks each signal independently with the same
method, it is much easy for us to discuss the security. However, current
exponential de Finetti theorem relying on the dimension and even diverges if
the dimension is infinite, while in practice the dimension is often unknown
or infinite.

There are also some other kind of quantum de Finetti theorems. In Ref. \cite%
{Matthias,Matthias2,finite de fi}, several de Finetti theorems for different
conditions are given. These de Finetti theorems can be independent with the
dimension. Even under the infinite dimensional case, they still converge.
However, These de Finetti theorems are polynomial and not exponential. As
the key size goes to infinite, they can not exponentially converge to zero.
Whether such polynomial de Finetti theorems can be applied to QKD requires
further discussion.

We can think about a more general case. Alice and Bob respectively get a
quantum state from a channel and do measurement and thus hold classical data
finally. Realistically, they only know the classical data and do not know
anything about the dimension of quantum state beforehand. Therefore, it is
not realistic for us to assume the dimension before discussing the security.
If the dimension of quantum state is unknown, current exponential de Finetti
theorem may not be directly applied and the security against the most
general attack is difficult to given by the information-theoretical method.
In Ref. \cite{Renner nature}, Renner gave some concrete examples to show the
de Finetti theorem. From these examples we can see that if the dimension of
individual quantum state is higher than the block size, the whole state may
be far away from an almost i.i.d. state. For some QKDs, the dimension
problem can be solved by introducing the squashing model \cite%
{squashingmode,squashingmode1}. However for some other protocols we do not
know whether there exist a squashing model, i.e. continuous variable (CV)
QKD \cite{binaryCVQKD,zhaoc} and differential phase shift (DPS) QKD \cite%
{DPSQKD}. Then, it is necessary for Alice and Bob to get some information
about their dimensions.

Here we give a general way to estimate the effective dimension (In
the following we will see that Alice and Bob's measurement data are
obtained almost only from a finite dimensional subspace. Here we
call the dimension of this subspace as effective dimension.) of a
system, and a general method to apply current information
theoretical security proof to practical QKD. Finally we prove that
if POVM elements corresponding to Alice and Bob's measured results
can be well described in a finite dimensional subspace with
sufficiently small error, the security of unknown dimensional system
is very close to that of a finite dimensional system, where Alice
and Bob put finite dimensional filters before their detectors. The
security of this finite dimensional system is covered by current
information theoretical security proof method. Then the security of
unknown-dimensional system can be solved. Our solution is based on
the estimation of the effective dimension of a system. In Ref.
\cite{min dimen}, Wehner et al. gave an estimation to the lower
bound of the dimension of a system. We hope future works can shrink
the gap between these two results. Up to now, some efforts have been
done for the finite key size case \cite{ZhaoIEEE,scar}. The security
under finite key size case may be much different from that under
infinite key size case. To give a better result for finite key size
case, it is necessary to give a tight estimation to the effective
dimension.

We may think that the world is always finite, so regard it as
guaranteed that current exponential de Finetti theorem can be
directly applied to practical system. It is not necessary the case.
Firstly, finite measurement result does not always mean finite
dimensional quantum state. A finite measurement result can also be
generated from an infinite quantum state. Secondly, to know the
upper bound of the dimension of quantum state is required if we
consider the finite key size case. From the Ref. \cite{Renner
thesis} we know that the amount of secret key rate under the finite
key size case is deeply related with the dimension. Our estimation
of effective dimension is expected to be favorable to finite key
size situation.

 We noted two parallel works shown in Ref. \cite{de
Finetti,lev}. In these
two works, the unconditional security of CVQKD is addressed. In Ref. \cite%
{de Finetti}, Renner et al. modified previous exponential de Finetti
theorem and this new theorem can be directly applied to CVQKD. From
this new de Finetti theorem we can see that in CVQKD if the variance
of Bob's measurement result is finite, the state Alice, Bob and Eve
share can still be approximated by an almost i.i.d. state. Our
result only works for heterodyne CVQKD and requires the maximum
value of Alice and Bob's heterodyne detection to be finite. Under
the infinite key size case, our result can give the same
approximation that the state describes the whole infinite
communications can be approximated by an almost product state with
arbitrarily small error. In Ref. \cite{lev}, Leverrier et al.
directly addressed the unconditional security of CVQKD without the
de Finetti theorem. Their work is based on the Gaussian optimality.
In this paper we approximate the CVQKD by a finite dimension
protocol. The security of finite dimension protocol can be covered
by current information theoretical security proof. Then the
unconditional security of CVQKD is possible to prove. Compared with
these two works, one advantage of our work is its application to
photon number detection protocols, e.g. another coherent state
protocol, DPSQKD. In the following, we will demonstrate how to apply
our result to DPSQKD.

The basic idea of our approach is as following. Although the dimension of
quantum state Alice, Bob and Eve initially share is totally unknown, after
obtaining measurement results, Alice and Bob collapse Eve's state into a
less complex state and can know some information about the effective
dimension of their state. Then we can construct another finite dimensional
protocol, where Alice and Bob put a finite dimensional filter right before
their detection equipments that can filter out high dimensional components.
We prove that final state of this new finite dimensional protocol is only
slightly different from the original one. Then the security of that
unknown-dimensional protocol can be approximated by this new protocol. The
security of this new finite dimensional protocol is covered by Ref. \cite%
{Renner thesis}, then the security of that unknown-dimensional protocol can
be solved.

In the following we will introduce a general QKD protocol at first
and then discuss unknown-dimensional problem. Latter we will
introduce a finite dimensional protocol and prove that if components
of POVM elements corresponding to Alice and Bob's measured results
on high dimensional bases are small enough then the security of
original unknown-dimensional protocol can be well approximated by
this finite dimensional protocol. While discussing their difference,
we will introduce an entanglement version measurement to describe
Alice and Bob's detection. Finally some application examples will be
given. In application examples, we will only discuss the infinite
key size case, while our result is also useful under the finite key
size case.

\section{Protocol:}

Here we limit our analysis to the following protocol.

Alice and Bob take $N$ quantum states from a channel respectively. Then they
permute their subsystem according to a commonly chosen random permutation.
They separate $N$ states into $l$ blocks and perform POVM measurement to
each state. Without loss of generality, we assume Alice and Bob respectively
hold several POVMs, $M^{Ai}=\{M_{x^{i}}^{Ai}\}$ and $M^{Bi}=\{M_{y^{i}}^{Bi}%
\}$ ($i=1,...,l$), where $x^{i}$ and $y^{i}$ denote corresponding
measurement results, and they perform the POVM $M^{Ai}=\{M_{x^{i}}^{Ai}\}$
and $M^{Bi}=\{M_{y^{i}}^{Bi}\}$ to the $i$-th blocks (we assume the choice
of POVMs is publicly known). Then they publish measurement results from the
first block to estimate the channel. Before the classical procedure they
estimate the dimension of their quantum state according to the region of
their measurement results. Then they give up partial of their measurement
results (required by the information-theoretical security proof \cite{Renner
thesis}) and finally obtain classical strings. After performing data
processing, information reconciliation and privacy amplification, they
finally generate secret keys. Here, we allow Alice and Bob to hold several
POVMs, mainly because in many QKD\ protocols, Alice and Bob need to randomly
change their measurement bases. One POVM corresponds to one choice of bases.

From current de Finetti theorem we know that if the dimension of the
channel is finite, the state Alice, Bob and Eve share after many
communications is close to an almost product state. It has been
shown that such almost product state almost has the same property
with the product state. The product state corresponds to the
collective attack. Then we only need to consider collective attack
\cite{Renner thesis}. However, if the dimension is infinite, the de
Finetti theorem may diverge. Then we cannot know the difference
between collective attack and coherent attack.

We assume after getting quantum state, Alice, Bob and Eve share the state $%
\rho _{A^{N}B^{N}E^{N}}$. Since discarding subsystem never increases mutual
information, we can safely assume that Eve holds the purification of $\rho
_{A^{N}B^{N}E^{N}}$, so that $\rho _{A^{N}B^{N}E^{N}}$ is pure \cite%
{Neilson,zhaoc,Renner thesis}. After measuring all $N$ states, Alice and Bob
know the region of their measurement results. For example, in DPSQKD \cite%
{DPSQKD}, if they use photon number resolving detector, they can know the
maximum photon number they received from one pulse. In CVQKD \cite%
{binaryCVQKD} with heterodyne detection, they can know the maximum amplitude
they get. Here, we will show that such information is enough for Alice and
Bob to know whether their system can be approximated by a finite dimensional
system.

Alice and Bob can make an initial estimation to their state
according to measurement results. After Alice and Bob knows the
region of their measurement results, they can only consider such
$\rho _{A^{N}B^{N}E^{N}}$ that can generate their measured results
with probability higher than certain small parameter $\varepsilon $.
Then the collection of states they need to consider is largely
reduced. The insecure probability introduced by such method is no
larger than $\varepsilon $ and the strength of security will be
reduced by $\varepsilon $ \cite{e-secure}. This procedure is
required by our proof.

More precisely, we assume Alice and Bob's measurement results from a
single state of $i$-th block belong to the region $\Xi (X^{i})$\ and
$\Xi (Y^{i})$ respectively. We let
\begin{equation}
D^{Ai}=\sum_{x^{i}\in \Xi (X^{i})}M_{x^{i}}^{Ai} \notag
\end{equation}
\begin{equation}
D^{Bi}=\sum_{y^{i}\in \Xi (Y^{i})}M_{y^{i}}^{Bi}  \notag
\end{equation}
Then $D^{Ai}$\ and $%
D^{Bi}$ actually are POVM elements that correspond to Alice and Bob's
measurement results belonging to the region $\Xi (X^{i})$\ and $\Xi (Y^{i})$
respectively. $D^{Ai}$ ($D^{Bi}$) may be different for different blocks. To
avoid distinguishing different $D^{Ai}$s ($D^{Bi}$s), here we define POVM
elements, $\tilde{D}^{A}$ and $\tilde{D}^{B}$ satisfying that for arbitrary
state $\rho $\ and $i$, we always have%
\begin{equation}
\begin{array}{c}
tr(\tilde{D}^{A}\rho )\geq tr(D^{Ai}\rho ) \\
tr(\tilde{D}^{B}\rho )\geq tr(D^{Bi}\rho )%
\end{array}%
\label{definition}
\end{equation}%
To know the requirement given in Eq. (\ref{definition}) well, we can see
some examples. It can be seen that $\tilde{D}^{A}=I$ is one trivial element
that always satisfy Eq. (\ref{definition}). Also, if all $D^{Ai}$s are just
the same, $\tilde{D}^{A}=D^{Ai}$ is the one satisfying Eq. (\ref{definition}%
). Furthermore, since for any $i$ and arbitrary $\rho $, the expectation
value of $\tilde{D}^{A}-D^{Ai}$ and $\tilde{D}^{B}-D^{Bi}$\ are
non-negative, $\tilde{D}^{A}-D^{Ai}$ and $\tilde{D}^{B}-D^{Bi}$\ are
non-negative operators. Therefore, if $\tilde{D}^{A}-D^{Ai}$ and $\tilde{D}%
^{B}-D^{Bi}$ are not zero, they are also valid POVM elements. Then $\{D^{Ai},%
\tilde{D}^{A}-D^{Ai},I-\tilde{D}^{A}\}$ and $\{D^{Bi},\tilde{D}^{B}-D^{Bi},I-%
\tilde{D}^{B}\}$ constitute a POVM respectively (It should be noted that $I$%
\ may be an operation of an infinite dimensional space.). Here we define the
POVM elements $\tilde{D}^{A}$ and $\tilde{D}^{B}$ mainly because the maximum
value of measurement result of different blocks may be different and then $%
D^{Ai}$s are not the same. Nevertheless, for most of current
protocols, it is not difficult to find a tight $\tilde{D}^{A}$ and
$\tilde{D}^{B}$. For example, in the heterodyne detection CVQKD,
Alice and Bob do not change the basis, so there are only two blocks,
one used for parameter estimation, one used to generate secret keys.
We assume at Alice's side the
maximum value of one block is $V_{A}^{\max 1}$, and that of the other is $%
V_{A}^{\max 2}$. Then $D^{A1}=\sum_{x^{i}\leq V_{A}^{\max 1}}M_{x^{i}}^{A1}$%
\ and $D^{A2}=\sum_{x^{i}\leq V_{A}^{\max 2}}M_{x^{i}}^{A2} $. Since Alice
uses the same POVM for these two blocks, we have $%
M_{x^{i}}^{A1}=M_{x^{i}}^{A2}$. If $V_{A}^{\max 1}>V_{A}^{\max 2}$, we can
choose $\tilde{D}^{A}=\sum_{x^{i}\leq V_{A}^{\max 1}}M_{x^{i}}^{A1}$, which
satisfies Eq. (\ref{definition}). Then $\tilde{D}^{A}-$ $D^{A1}=0$ and $%
\tilde{D}^{A}-$ $D^{A2}=\sum_{V_{A}^{\max 2}<x^{i}\leq V_{A}^{\max
1}}M_{x^{i}}^{A1}$, which is a POVM element. In the DPSQKD, Bob also does
not change his bases, so a similar result can be obtained. \

To analysis the security we can only consider the state $\rho
_{A^{N}B^{N}E^{N}}$ that satisfies
\begin{equation}
tr[(\tilde{D}^{A}\tilde{D}^{B})^{\otimes N}\rho _{A^{N}B^{N}E^{N}}]\geq
\varepsilon  \label{requirement}
\end{equation}%
while the final strength of security will be reduced by $\varepsilon $,
where $\tilde{D}^{A}$ and $\tilde{D}^{B}$ are properly chosen elements that
satisfy Eq. (\ref{definition}). Then the collection of $\rho
_{A^{N}B^{N}E^{N}}$\ we need to consider is largely reduced.\ It can be seen
that to shrink the collection of $\rho _{A^{N}B^{N}E^{N}}$, we need to find
tight $\tilde{D}^{A}$ and $\tilde{D}^{B}$. In the following, we will see
that this technique is required by our argument.

After Alice and Bob's measurement, the state Alice, Bob and Eve hold becomes
$\rho _{X^{N}Y^{N}E^{N}}$, where $X$ and $Y$\ are classical variable that
can take the value $x^{i}$ and $y^{i}$ and can be expressed by orthogonal
quantum state \cite{Renner thesis}. Actually, the security of QKD system
directly related to the state $\rho _{X^{N}Y^{N}E^{N}}$, rather than
original state $\rho _{A^{N}B^{N}E^{N}}$. Therefore, if we can find a finite
dimensional system that generates another state $\tilde{\rho}%
_{X^{N}Y^{N}E^{N}}$ very close to $\rho _{X^{N}Y^{N}E^{N}} $, then the
security of the original unknown system can be approximated by this finite
dimensional system.

\begin{figure}[tbp]
\includegraphics[width=8cm]{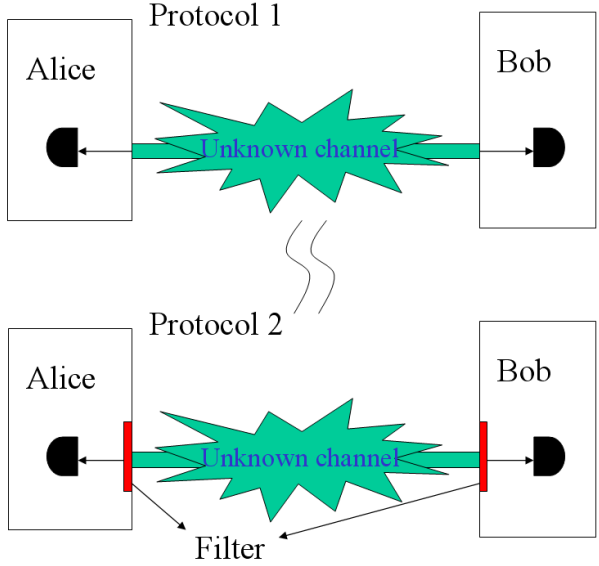}
\caption{Illustration of protocol 1 and protocol 2, where in
protocol 2 finite dimensional filters are put before the detectors.
If the filters are properly chosen, the security of protocol 1 can
be well approximated by that of protocol 2, while the security of
protocol 2 is covered by current information theoretical security
proof method. The exact difference between protocol 1 and protocol 2
becomes significant while we consider the the finite key size case.}
\label{fig2}
\end{figure}

Now we can compare two schemes as illustrated in Fig. \ref{fig2}. One is the
original unknown-dimensional scheme, and the other is a modified scheme, in
which Alice and Bob respectively put filters before their detectors. We
assume these two filters can totally filter out high dimensional component
of received state and dimensions of output states of these two filters are $%
d_{A}$ and $d_{B}$ respectively. For convenience, we will call the original
protocol as protocol 1 and the modified one as protocol 2. Then in the
protocol 2 dimensions of Alice and Bob's received states are $d_{A}$ and $%
d_{B}$ respectively. In the following, we will see that if we properly set
the filter and choose high enough $d_{A}$ and $d_{B}$, then the security of
protocol 1 can be approximated by that of protocol 2.

To simplify our discussion, it is necessary to avoid distinguishing
different blocks. We know that $\tilde{D}^{A}-D^{Ai}$ and $\tilde{D}%
^{B}-D^{Bi}$\ are also POVM elements. Here we introduce other two classical
data $x^{\prime i}$ and $y^{\prime i}$ that correspond to POVM elements $%
\tilde{D}^{A}-D^{Ai}$ and $\tilde{D}^{B}-D^{Bi}$ respectively. Then in
protocol 1 Alice and Bob's measurement results of $i$-th block are within
the region $\Xi (X^{i})\cup \{x^{\prime i}\}$\ and $\Xi (Y^{i})\cup
\{y^{\prime i}\}$ respectively. Therefore, this protocol does not change if
it runs as follows. While getting measurement results from a state of $i$-th
block, Alice and Bob accept them only when they belong to region $\Xi
(X^{i})\cup \{x^{\prime i}\}$ and $\Xi (Y^{i})\cup \{y^{\prime i}\}$
respectively. Otherwise, they discard them. Now we can calculate the
difference between the protocol 1 and the protocol 2.

\section{Estimation of $L_{1}$-distance based on observations:}

\begin{figure}[tbp]
\includegraphics[width=8cm]{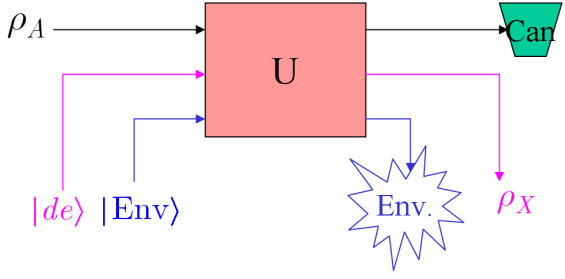}
\caption{Illustration of entanglement version measurement, where $|$Env$%
\rangle$ and $|de\rangle$ respectively denote the initial state of
environment and the detector. The measurement can be realized by the unitary
operation among received state, detector and the environment. After the
operation, the received state and the environment are given up and the
measurement result can be directly given by the state of detector, denoted
by $\protect\rho_{X}$.}
\label{fig1}
\end{figure}

Before calculating the difference, here we introduce a entanglement version
measurement. There are several interpretations for the quantum measurement,
e.g. von Neumann measurement scheme and Many-worlds interpretation \cite%
{quantum measure}. Here we are not to give a new philosophical
interpretation, but to construct a physical model that can effectively
perform POVM measurement. This physical model allows us to easily find the
difference between protocol 1 and protocol 2. For briefness, here we only
take Alice's measurement as an example. Alice's POVM measurement can be
performed by the equipment shown in Fig. \ref{fig1}. The measurement
procedure is realized by an interaction among her received state, detector
and the environment. After the interaction, Alice gives up received state
and the environment and thus only holds the detector, which directly gives
her the classical data. After reading out the classical data, Alice set the
detector and environment to the initial state to do the next measurement. In
this model we require initial states of Alice's detector and environment are
pure and respectively to be $|de\rangle _{A}$ and $|$Env$\rangle _{A}$. For
convenience, here we let $|ini\rangle _{A}$\ denote $|de\rangle _{A}|$Env$%
\rangle _{A}$. Then the interaction among the received state, detector and
environment for $i$-th block can be given by%
\begin{equation}
U_{A}^{i}=\sum_{x^{i}}|x^{i}\rangle |Q_{x^{i}}\rangle _{A}\langle ini|\sqrt{%
M_{x^{i}}^{Ai}}  \label{unitary}
\end{equation}%
where $|x^{i}\rangle $s are orthogonal states of detector, $%
|Q_{x^{i}}\rangle $ describes orthogonal state of the environment, $\langle
ini|$ denotes the initial pure state of Alice's detector and environment and
$\sqrt{M_{x^{i}}^{Ai}}$ is the POVM operators corresponding to POVM element $%
M_{x^{i}}^{Ai}$ \cite{Neilson}. To check the validity of this measurement,
we can apply it to a two parties system $\rho _{AB}$. After the interaction
described by $U_{A}^{i}$, the state of whole system becomes
\begin{eqnarray*}
\rho _{ABXQ_{X}} &=&U_{A}^{i}\rho _{AB}\otimes |ini\rangle _{A}\langle
ini|U_{A}^{i+} \\
&=&\sum_{x^{i}}|x^{i}\rangle |Q_{x^{i}}\rangle \sqrt{M_{x^{i}}^{Ai+}}\rho
_{AB}\sum_{x^{j}}\langle x^{j}|\langle Q_{x^{j}}|\sqrt{M_{x^{j}}^{Aj}}
\end{eqnarray*}%
where $Q_{X}$ denotes the environment and all of $|Q_{x^{i}}\rangle $s are
orthogonal with each other. After we trace out the system A and environment
we immediately obtain the state
\begin{equation*}
\rho _{XB}=\sum_{x^{i}}|x^{i}\rangle \langle x^{i}|\otimes tr_{A}(\sqrt{%
M_{x^{i}}^{Ai+}}\rho _{AB}\sqrt{M_{x^{i}}^{Ai}})
\end{equation*}%
We see that
\begin{equation*}
tr_{A}(\sqrt{M_{x^{i}}^{Ai+}}\rho _{AB}\sqrt{M_{x^{i}}^{Ai}})=P(x^{i})\rho
_{B}^{x^{i}}
\end{equation*}%
where $P(x^{i})$ is the probability of the out come $x^{i}$ and $\rho
_{B}^{x^{i}}$ denotes Bob's conditional state while Alice's measurement
result is $x^{i}$. Then $\rho _{XB}$\ becomes
\begin{equation*}
\rho _{XB}=\sum_{x^{i}}P(x^{i})|x^{i}\rangle \langle x^{i}|\otimes \rho
_{B}^{x^{i}}
\end{equation*}%
which consists with the POVM measurement. Since Alice only accept the data
within the collection $\Xi (X^{i})\cup \{x^{\prime i}\}$, we can reduce the
unitary transformation given in Eq. \ref{unitary} to a general quantum
operation $\hat{O}_{A}^{i}$ to describe Alice's effective measurement, which
is given by
\begin{equation}
\hat{O}_{A}^{i}=\sum_{x^{i}\in \Xi (X^{i})\cup \{x^{\prime
i}\}}|x^{i}\rangle |Q_{x^{i}}\rangle _{A}\langle ini|\sqrt{M_{x^{i}}^{Ai}}
\label{Osin}
\end{equation}%
Here we can see that $\hat{O}_{A}^{i}$\ may not be a unitary transformation.
The quantum operation that describes Alice's total $N$ detection is%
\begin{equation}
\hat{O}_{A^{N}}=\bigotimes\limits_{i=1}^{l}(\hat{O}_{A}^{i})^{\otimes n_{i}}
\label{Ototal}
\end{equation}%
where $n_{i}$ denotes the length of $i$-th block. By the same way we can
give the operator describing Bob's whole $N$ detections (By substituting the
notation $A$ by $B$.).

Since we only need to consider the case that $\rho _{A^{N}B^{N}E^{N}}$ is
pure, here we assume the initial state Alice and Bob receive is $|\Psi
_{A^{N}B^{N}E^{N}}\rangle $. The initial state of Alice and Bob's detector
and environment is $|\Psi _{X^{N}Y^{N}Q_{X}^{N}Q_{Y}^{N}}\rangle
=|ini\rangle _{A}|ini\rangle _{B}$. Then for the protocol-1, after Alice and
Bob's measurement (general quantum operation \cite{Neilson}) the state
describing Alice, Bob, Eve, detectors and environment becomes
\begin{eqnarray}
|\Psi _{P-1}\rangle &=&|\Psi
_{X^{N}Y^{N}A^{N}B^{N}Q_{X}^{N}Q_{Y}^{N}E^{N}}\rangle  \notag \\
&=&\frac{\hat{O}_{A^{N}}\hat{O}_{B^{N}}|\Psi _{A^{N}B^{N}E^{N}}\rangle |\Psi
_{X^{N}Y^{N}Q_{X}^{N}Q_{Y}^{N}}\rangle }{\sqrt{\langle \hat{O}_{A^{N}}^{+}%
\hat{O}_{B^{N}}^{+}\hat{O}_{A^{N}}\hat{O}_{B^{N}}\rangle }}  \label{state p1}
\end{eqnarray}%
where $X$ and $Y$\ denotes Alice and Bob's detectors, $Q_{X}$ and $Q_{Y}$
denote the environment around Alice and Bob, $\hat{O}_{B^{N}}$ is the
operator describing Bob's whole $N$ detections and $\langle \hat{O}%
_{A^{N}}^{+}\hat{O}_{B^{N}}^{+}\hat{O}_{A^{N}}\hat{O}_{B^{N}}\rangle $
describes expectation value of $\hat{O}_{A^{N}}^{+}\hat{O}_{B^{N}}^{+}\hat{O}%
_{A^{N}}\hat{O}_{B^{N}}$.

In protocol-2 Alice and Bob respectively put a $d_{A}$ and $d_{B}$
dimensional filter before their detectors. The filter can be described by a
projection into a subspace. Here we let the projector $P_{A}^{d_{A}}$ and $%
P_{B}^{d_{B}}$ denotes Alice and Bob's filters. For convenience we let $%
P_{AB}^{d_{A}d_{B}}$ denotes $P_{A}^{d_{A}}\otimes P_{B}^{d_{B}}$. Then for
the protocol-2 after Alice and Bob's measurement the whole state becomes%
\begin{gather}
|\Psi _{P-2}\rangle =|\tilde{\Psi}%
_{X^{N}Y^{N}A^{N}B^{N}Q_{X}^{N}Q_{Y}^{N}E^{N}}\rangle  \label{state p2} \\
=\frac{\hat{O}_{A^{N}}\hat{O}_{B^{N}}(P_{AB}^{d_{A}d_{B}})^{\otimes N}|\Psi
_{A^{N}B^{N}E^{N}}\rangle |\Psi _{X^{N}Y^{N}Q_{X}^{N}Q_{Y}^{N}}\rangle }{%
\sqrt{\langle (P_{AB}^{d_{A}d_{B}})^{\otimes N}\hat{O}_{A^{N}}^{+}\hat{O}%
_{B^{N}}^{+}\hat{O}_{A^{N}}\hat{O}_{B^{N}}(P_{AB}^{d_{A}d_{B}})^{\otimes
N}\rangle }}  \notag
\end{gather}

After tracing out received quantum state $A^{N}$ and $B^{N}$, and
the environment $Q^{N}_{X}$ and $Q^{N}_{Y}$ we obtain the state
Alice, Bob and Eve finally hold. For the protocol-1, they finally
hold the state $\rho
_{X^{N}Y^{N}E^{N}}$ and for the protocol-2, they finally share $\tilde{\rho}%
_{X^{N}Y^{N}E^{N}}$. If we know the $L_{1}$ distance between $\rho
_{X^{N}Y^{N}E^{N}}$ and $\tilde{\rho}_{X^{N}Y^{N}E^{N}}$ we can know the
difference between securities of protocol-1 and protocol-2 \cite{Renner
thesis}. Since tracing out the subsystem never increases the $L_{1}$
distance \cite{Renner thesis}, the $L_{1}$\ distance between $\rho
_{X^{N}Y^{N}E^{N}}$ and $\tilde{\rho}_{X^{N}Y^{N}E^{N}}$ is no larger than
that between $|\Psi _{P-1}\rangle $ and $|\Psi _{P-2}\rangle $.\ We know
that
\begin{gather}
|\Psi _{A^{N}B^{N}E^{N}}\rangle =  \label{decomposition} \\
(P_{AB}^{d_{A}d_{B}})^{\otimes N}|\Psi _{A^{N}B^{N}E^{N}}\rangle +\overline{%
(P_{AB}^{d_{A}d_{B}})^{\otimes N}}|\Psi _{A^{N}B^{N}E^{N}}\rangle  \notag
\end{gather}%
where $\overline{P}$ denotes the orthogonal complement space of $P$. By
putting the Eq. (\ref{decomposition}) into Eq. (\ref{state p1}) we quickly
know
\begin{equation*}
|\Psi _{P-1}\rangle =\alpha |\Psi _{P-2}\rangle +\beta |\Psi ^{\prime
}\rangle
\end{equation*}%
where
\begin{equation}
|\beta |=\frac{\sqrt{\langle \overline{(P_{AB}^{d_{A}d_{B}})^{\otimes N}}%
\hat{O}_{A^{N}}^{+}\hat{O}_{B^{N}}^{+}\hat{O}_{A^{N}}\hat{O}_{B^{N}}%
\overline{(P_{AB}^{d_{A}d_{B}})^{\otimes N}}\rangle }}{\sqrt{\langle \hat{O}%
_{A^{N}}^{+}\hat{O}_{B^{N}}^{+}\hat{O}_{A^{N}}\hat{O}_{B^{N}}\rangle }}.
\label{beta}
\end{equation}%
and
\begin{equation*}
|\Psi ^{\prime }\rangle =\frac{\hat{O}_{A^{N}}\hat{O}_{B^{N}}\overline{%
(P_{AB}^{d_{A}d_{B}})^{\otimes N}}|\Psi _{A^{N}B^{N}E^{N}}\rangle |\Psi
_{X^{N}Y^{N}Q_{X}^{N}Q_{Y}^{N}}\rangle }{\sqrt{\langle \overline{%
(P_{AB}^{d_{A}d_{B}})^{\otimes N}}\hat{O}_{A^{N}}^{+}\hat{O}_{B^{N}}^{+}\hat{%
O}_{A^{N}}\hat{O}_{B^{N}}\overline{(P_{AB}^{d_{A}d_{B}})^{\otimes N}}\rangle
}}
\end{equation*}%
is a state obtained from the complimentary space $\overline{%
(P_{AB}^{d_{A}d_{B}})^{\otimes N}}$, which may not be orthogonal with $|\Psi
_{P-2}\rangle $. From the Appendix-A of Ref. \cite{Renner thesis} we know
that the $L_{1}$\ distance of two pure state $|\Psi _{1}\rangle $ and $|\Psi
_{2}\rangle $\ can be given by
\begin{equation*}
|||\Psi _{1}\rangle -|\Psi _{2}\rangle ||=2\sqrt{1-|\langle \Psi _{1}|\Psi
_{2}\rangle |^{2}}
\end{equation*}%
where $||\cdot ||$ denotes the $L_{1}$ distance. Then the $L_{1}$ distance
between $|\Psi _{P-1}\rangle $ and $|\Psi _{P-2}\rangle $ is no larger than $%
2|\beta |$, which yields
\begin{equation*}
||\rho _{X^{N}Y^{N}E^{N}}-\tilde{\rho}_{X^{N}Y^{N}E^{N}}||\leq 2|\beta |
\end{equation*}

For convenience here we let $\tilde{D}=\tilde{D}^{A}\tilde{D}^{B}$. If we
put Eqs. (\ref{Osin}) and (\ref{Ototal}) into Eq. (\ref{beta}) and apply the
fact that operators of different detectors are commutate, we can know that
\begin{equation}
|\beta |=\sqrt{\frac{tr_{ABE}[\tilde{D}^{\otimes N}\overline{%
(P_{AB}^{d_{A},d_{B}})^{\otimes N}}\rho _{A^{N}B^{N}E^{N}}\overline{%
(P_{AB}^{d_{A},d_{B}})^{\otimes N}}]}{tr_{ABE}[\tilde{D}^{\otimes N}\rho
_{A^{N}B^{N}E^{N}}]}}  \label{beta2}
\end{equation}%
Now we can see that if all of pure state $\rho _{A^{N}B^{N}E^{N}}$
satisfying Eq. (\ref{requirement}) make $|\beta |$ small enough, then
protocol-1 can be well approximated by protocol-2.

To estimate $|\beta |$ here we give a very useful theorem.

\begin{theorem}
Let $|1\rangle _{A}$, $|2\rangle _{A}$, ..., and $|1\rangle _{B}$, $%
|2\rangle _{B}$, ..., be bases of Alice and Bob's Hilbert spaces
respectively, by which projectors $P_{A}^{d_{A}}$ and $P_{B}^{d_{B}}$ can be
respectively given by $P_{A}^{d_{A}}=|1\rangle _{A}\langle
1|+...+|d_{A}\rangle _{A}\langle d_{A}|$ and $P_{B}^{d_{B}}=|1\rangle
_{B}\langle 1|+...+|d_{B}\rangle _{B}\langle d_{B}|$. Then if we have $%
\sum_{i=1,j=d_{A}}^{\infty ,\infty }|_{A}\langle i|\tilde{D}^{A}|j\rangle
_{A}|+\sum_{i=1,j=d_{B}}^{\infty ,\infty }|_{B}\langle i|\tilde{D}%
^{B}|j\rangle _{B}|\leq $\ $\frac{\varepsilon ^{3}}{N}$, for arbitrary $\rho
_{A^{N}B^{N}E^{N}}$ it is always satisfied that $tr_{AB}[\tilde{D}^{\otimes
N}\overline{(P_{AB}^{d_{A}d_{B}})^{\otimes N}}\rho _{A^{N}B^{N}}\overline{%
(P_{AB}^{d_{A}d_{B}})^{\otimes N}}]:=L\leq \varepsilon ^{3}$.
\end{theorem}

\textit{Proof}: We can see that $L$\ is no larger than $\max_{|\Psi
^{N}\rangle }\langle \Psi ^{N}|\tilde{D}^{\otimes N}|\Psi ^{N}\rangle $,
where $|\Psi ^{N}\rangle \in \overline{(P_{AB}^{d_{A}d_{B}})^{\otimes N}}$.
If we expand $\langle \Psi ^{N}|\tilde{D}^{\otimes N}|\Psi ^{N}\rangle
=\langle \Psi ^{N}|\overline{(P_{AB}^{d_{A}d_{B}})^{\otimes N}}\tilde{D}%
^{\otimes N}|\Psi ^{N}\rangle $ into product spaces $\overline{P_{A}^{d_{A}}}%
(P_{A}^{d_{A}})^{\otimes N-1}(P_{B}^{d_{B}})^{\otimes N},...,$ and do
straightforward calculation, we can immediately find that $L\leq
N[\sum_{i=1,j=d_{A}}^{\infty ,\infty }|_{A}\langle i|\tilde{D}^{A}|j\rangle
_{A}|+\sum_{i=1,j=d_{B}}^{\infty ,\infty }|_{B}\langle i|\tilde{D}%
^{B}|j\rangle _{B}|]\leq $\ $\varepsilon ^{3}$. (The straightforward
calculation is too bothering to show here. Detailed one can be seen in the
appendix.) \ \ \ \ \ $\square $

Since $tr_{ABE}[\tilde{D}^{\otimes N}\rho _{A^{N}B^{N}E^{N}}]\geq
\varepsilon $, from Eq. (\ref{beta2}), we can see that Theorem 1 actually
gives a sufficient condition for $|\beta |\leq \varepsilon$.

Now, we can know the distance between protocol 1 and protocol 2 from the
measurement results. The only remained problem is to give the difference
between securities of protocol-1 and protocol-2 if the state difference of
them is known.

\begin{theorem}
If $||\rho _{X^{N}Y^{N}E^{N}}-\tilde{\rho}_{X^{N}Y^{N}E^{N}}||\leq 2 \delta $
for all $\rho _{A^{N}B^{N}E^{N}}$ satisfying $tr[(\tilde{D}^{A}\tilde{D}%
^{B})^{\otimes N}\rho _{A^{N}B^{N}E^{N}}]\geq \delta $, then the $5\delta
+\epsilon $-secure secret key rate of protocol-1 is no less than the $%
2\delta+\epsilon $-secure secret key rate of protocol-2, while Alice
and Bob take results from protocol-1 as that from protocol-2 to
estimate the secret key rate of protocol-2 by the information
theoretical method.
\end{theorem}

\textit{Proof}: The $L_{1}$ distance cannot be increased by quantum
operations and thus classical bit-wise processing \cite{Renner thesis}. If $%
||\rho _{X^{N}Y^{N}E^{N}}-\tilde{\rho}_{X^{N}Y^{N}E^{N}}||\leq 2\delta $,
then we have $||\rho _{\bar{X}^{k}\bar{E}}-\tilde{\rho}_{\bar{X}^{k}\bar{E}%
}||\leq 2\delta $ and $||\rho _{X^{N}Y^{N}}-\tilde{\rho}_{X^{N}Y^{N}}||\leq
2\delta $, where $\bar{X}^{k}$, $\bar{Y}^{k}$ and $\bar{E}$ denote Alice and
Bob's classical data and Eve's state after the data processing respectively,
during which some information maybe announced. The security is well defined
by smooth min- and max-entropies. The amount of $\epsilon $-secure secret
keys can be given by $H_{\min }^{\epsilon ^{\prime }}(\rho _{\bar{X}^{k}\bar{%
E}}|\bar{E})-leak_{IR}$ \cite{expl}, while the strength of parameter
estimation is $\epsilon ^{\prime \prime \prime }$, where $H_{\min
}^{\epsilon ^{\prime }}(\cdot |\cdot )$ denotes the smooth min-entropy, $%
leak_{IR}$ denotes the amount of information published during the $\epsilon
^{\prime \prime }$-secure reconciliation and $\epsilon ^{\prime }+\epsilon
^{\prime \prime }+\epsilon ^{\prime \prime \prime }=\epsilon $ \cite{Renner
thesis}. Since $||\rho _{\bar{X}^{k}\bar{E}}-\tilde{\rho}_{\bar{X}^{k}\bar{E}%
}||\leq 2\delta $, the smooth min-entropy satisfies $H_{\min }^{2\delta
+\epsilon }(\rho _{\bar{X}^{k}\bar{E}}|\bar{E})\geq H_{\min }^{\epsilon }(%
\tilde{\rho}_{\bar{X}^{k}\bar{E}}|\bar{E})$ \cite{Renner thesis}. Also, if
Alice and Bob use the data from the protocol-1 as that from the protocol-2
to estimate the state of protocol-2, the security of the parameter
estimation \cite{Renner thesis} will be reduced by $2\delta $, because $%
||\rho _{X^{N}Y^{N}}-\tilde{\rho}_{X^{N}Y^{N}}||\leq 2\delta $. Furthermore,
if we only consider the $\rho _{A^{N}B^{N}E^{N}}$ satisfying $tr[(\tilde{D}%
^{A}\tilde{D}^{B})^{\otimes N}\rho _{A^{N}B^{N}E^{N}}]\geq \delta $, the
strength of security will also be reduced by $\delta $. In all, the $%
\epsilon ^{\prime }+\epsilon ^{\prime \prime }+\epsilon ^{\prime \prime
\prime }$ secure security of protocol 2 is given by $H_{\min }^{\epsilon
^{\prime }}(\tilde{\rho}_{\bar{X}^{k}\bar{E}}^{{}}|\bar{E})-leak_{IR}$,
while the strength of parameter estimation is $\epsilon ^{\prime \prime
\prime }$\ and the $\tilde{\rho}_{\bar{X}^{k}\bar{E}}$ is estimated by the
data obtained from protocol 2. The $\epsilon ^{\prime }+\epsilon ^{\prime
\prime }+\epsilon ^{\prime \prime \prime }+2\delta $ secure security of
protocol 2 is given by $H_{\min }^{\epsilon ^{\prime }}(\tilde{\rho}_{\bar{X}%
^{k}\bar{E}}^{{}}|\bar{E})-leak_{IR}$, while the strength of parameter
estimation is $\epsilon ^{\prime \prime \prime }+2\delta $, where the $%
\tilde{\rho}_{\bar{X}^{k}\bar{E}}^{{}}$ is estimated by data obtained from
protocol 1. Finally the $\epsilon ^{\prime }+\epsilon ^{\prime \prime
}+\epsilon ^{\prime \prime \prime }+5\delta $ secure secret key rate of
protocol 1 can be given by $H_{\min }^{2\delta +\epsilon ^{\prime }}(\rho _{%
\bar{X}^{k}\bar{E}}|\bar{E})-leak_{IR}\geq H_{\min }^{\epsilon ^{\prime }}(%
\tilde{\rho}_{\bar{X}^{k}\bar{E}}^{{}}|\bar{E})-leak_{IR}$, while the
strength of parameter estimation is $\epsilon ^{\prime \prime \prime
}+2\delta $, the state $\tilde{\rho}_{\bar{X}^{k}\bar{E}}^{{}}$ is estimated
by the data obtained from protocol 1 and only the $\rho _{A^{N}B^{N}E^{N}}$
satisfying $tr[(\tilde{D}^{A}\tilde{D}^{B})^{\otimes N}\rho
_{A^{N}B^{N}E^{N}}]\geq \delta $ is considered. Here the term $H_{\min
}^{\epsilon ^{\prime }}(\tilde{\rho}_{\bar{X}^{k}\bar{E}}^{{}}|\bar{E}%
)-leak_{IR}$ is amount of the $\epsilon ^{\prime }+\epsilon ^{\prime
\prime }+\epsilon ^{\prime \prime \prime }+2\delta $ secure secrete
keys of protocol 2, while the strength of parameter estimation is
$\epsilon ^{\prime \prime \prime }+2\delta $, where $2\delta $ comes
from the fact that the state
$\tilde{\rho}_{\bar{X}^{k}\bar{E}}^{{}}$ is estimated by the data
obtained from protocol 1.\ \ \ \ \ \ \ \ \ \ \ $\square $ \ \

The state distance $||\rho _{X^{N}Y^{N}E^{N}}-\tilde{\rho}%
_{X^{N}Y^{N}E^{N}}||\leq 2\delta $\ can be evaluated from the
measurement results through theorem 1. The security of protocol 2 is
covered by current information theoretical security proof method.
Then the the security of protocol 1 can be solved. \

\section{Security of Protocol 2.}

If Alice and Bob's received initial state in protocol 1 is $\rho
_{A^{N}B^{N}}$, the state they received in protocol 2 is $\tilde{\rho}%
_{A^{N}B^{N}}=\frac{1}{p}(P_{AB}^{d_{A}d_{B}})^{\otimes N}\rho
_{A^{N}B^{N}}(P_{AB}^{d_{A}d_{B}})^{\otimes N}$, where $\frac{1}{p}$ is
introduced for normalization. After quantum communication, Alice and Bob
will permute their state, then $\rho _{A^{N}B^{N}}$\ is permutation
invariant. The projection operator $(P_{AB}^{d_{A}d_{B}})^{\otimes N}$\
commutates with the permutation operator, so $\tilde{\rho}_{A^{N}B^{N}}$ is
also a permutation invariant state. The dimension of individual state of $%
\tilde{\rho}_{A^{N}B^{N}}$ is $d_{A}d_{B}$. Then there is a symmetric
purification for $\tilde{\rho}_{A^{N}B^{N}}$ in a Hilbert space of dimension
$(d_{A}d_{B})^{2N}$, which actually is $\tilde{\rho}_{A^{N}B^{N}E^{N}}$ \cite%
{Renner thesis,expl4}. Then the dimension of the individual state\ of $\tilde{\rho}%
_{A^{N}B^{N}E^{N}}$\ is $(d_{A}d_{B})^{2}$. According to current
exponential de Finetti theorem, the state
$\tilde{\rho}_{A^{N}B^{N}E^{N}}$\ is close to an almost product
state \cite{expl5}. Then we can only consider the collective attack.
Since under the collective attack, Eve attacks all of signals
independently by the same method, here we let $\tilde{\rho}_{ABE}$
denotes the state Alice, Bob and Eve share after a single
communication. Before calculating the secret key rate, we need to
estimate possible $\tilde{\rho}_{ABE}$\ from measurement results. It
should be noted that although $\tilde{\rho}_{ABE}$\ belongs to a
Hilbert space of dimension $(d_{A}d_{B})^{2}$, we do not really need
to estimate it only in a $(d_{A}d_{B})^{2}$ dimensional subspace. We
can still construct it in an infinite dimensional space, because a
state belonging to a $(d_{A}d_{B})^{2}$ dimensional Hilbert space
also belongs to a infinite dimensional Hilbert space \cite{expl2}.
This point shows that while we discuss the collective attack for
protocol 2, we do not need to take the filter in to account. If we
give up filters in protocol 2, the protocol 2 becomes the same as
protocol 1. Then if we do not take the filter into account, the
security against collective attack of protocol 2 is actually
equivalent to that of protocol 1. Finally, our conclusion is as
follows. The security of protocol 1 can be approximated by that of
protocol 2. For the protocol 2 we only need to consider the
collective attack. While the Hilbert space of protocol 2 is only a
subspace of protocol 1, then the secrete key rate of protocol 2
against collective attack is no less than that of protocol 1 against
collective attack. Finally, we actually give the difference between
coherent attack and collective attack for protocol 1. We introduce
the filter only to apply current de Finetti theorem and to give the
difference between coherent attack and collective attack for
protocol 1.

The $5\delta +\epsilon $-secure unconditional secret key rate of
protocol-1 is no less than the $2\delta+\epsilon $-secure secret key
rate of protocol-2. Under the infinite key size case, the
unconditional secrete key rate of protocol 2 is given by the secret
key rate against collective attacks \cite{Renner thesis}. The secret
key rate under collective attack of protocol 2 is no less than that
of protocol 1. Also under the infinite key size case, the parameter
$\epsilon $ can approach to zero. Then the $5\delta $-secure
unconditional secret key rate of protocol-1 is no less than the
$2\delta$ secure unconditional secret key rate of protocol-2 and no
less than its secret key rate against collective attacks, where
$2\delta$ comes from the fact that Alice and Bob use the data of
protocol 1 to estimate the state of protocol 2. In addition, under
the infinite key size
case, we may  choose large enough $d_{A}$ and $d_{B}$ so as to make $\delta $%
approach to zero. Then we can directly say that for protocol 1 if
the POVM elements corresponding to the measured results can be
arbitrarily well described in a finite dimensional space, the
collective attack is optimal under the infinite key size case.

For many practical QKDs, the projection of POVM elements of measured results
on high dimensional basis is extremely small. For example, the POVM element
for heterodyne detection corresponding to measured result $(p,q)$ is $%
M_{p,q}=\frac{1}{\pi }|p+iq\rangle \langle p+iq|$, whose component on the
photon number basis $|m\rangle $ exponentially goes to zero as $m$ increase.
The POVM of inefficient photon number resolving detector \cite{POVM} also
has similar property. Then if a QKD protocol utilize such detectors, Alice
and Bob can announce the maximum $p^{2}+q^{2}$ or maximum photon number
received from one pulse. Then Alice and Bob can construct the big POVM $%
\tilde{D}$ and for a given\ $\frac{\varepsilon ^{3}}{N}$ they can find a big
enough $d$ (smaller than $N$) that in photon number picture satisfies $%
\sum_{i=1,j=d}^{\infty ,\infty }|\langle i|\tilde{D}|j\rangle |\leq $\ $%
\frac{\varepsilon ^{3}}{N}$. Then the difference between states of
protocol 1 and protocol 2 can be smaller than $2\varepsilon $. The $%
5\varepsilon +\epsilon $-secure secret key rate can be given by $2\varepsilon+\epsilon $%
-secure secret key rate of protocol 2, which is covered by Ref.
\cite{Renner thesis}.

\section{Applications:}

In the realistic case, the measured result is always finite. In
heterodyne detection protocols, the maximum value of measured result
is limited. In photon number detection protocol, the maximum
received photon number is finite. Such realistic cases allow us
readily apply our results.

Here we give two application examples. We will see that our result
can be readily used for heterodyne detection and photon number
detection case. It should be noted that in the following we only
proved that for CVQKD and DPSQKD the collective attack is optimal
under infinite key size case. How to prove their security against
collective attack has not been solved in this paper. For short, we
only take the infinite key size case for examples. It seems that our
estimation of effective dimension is meaningless under this case.
However, we should note that under the finite key size case, the
estimation of effective dimension will be useful.

\subsection{Unconditional security of CVQKD}

Now we apply our results to the heterodyne detection CVQKD and prove
that as the key size goes to infinite the collective is optimal. In
the prepare \& measurement CVQKD, Alice prepare a continuous
variable EPR pair, and sends one part to Bob. Alice and Bob
respectively do heterodyne detection to their held states. The
security of such scheme against collective attack is discussed in
Ref. \cite{binaryCVQKD}. Here, we prove that for this protocol the
collective attack is optimal under the
infinite key size case. We denote Alice and Bob's measurement result by $%
(p_{A},q_{A})$ and $(p_{B},q_{B})$ respectively. The corresponding POVM
elements are respectively $M_{p_{A},q_{A}}=\frac{1}{\pi }|p_{A}+iq_{A}%
\rangle \langle p_{A}+iq_{A}|$ and $M_{p_{B},q_{B}}=\frac{1}{\pi }%
|p_{B}+iq_{B}\rangle \langle p_{B}+iq_{B}|$. In a realistic system, the
maximum value of Alice and Bob's measurement results is finite (or Alice and
Bob can give up some extremely larger measurement results). Then their final
shared data is within certain region. We assume $V_{A}^{\max }$ and $%
V_{B}^{\max }$ are large enough, so that for all possible $(p_{A},q_{A})$s
and $(p_{B},q_{B})$s Alice and Bob hold satisfy $p_{A}^{2}+q_{A}^{2}\leq
V_{A}^{\max }$ and $p_{B}^{2}+q_{B}^{2}\leq V_{B}^{\max }$ (or Alice and Bob
only accept the data with amplitude no larger than $V_{A}^{\max }$ and $%
V_{B}^{\max }$). Then we can construct $\tilde{D}^{A}$ and $\tilde{D}^{B}$
respectively to be
\begin{eqnarray*}
\tilde{D}^{A} &=&\frac{1}{\pi }\int\limits_{p_{A}^{2}+q_{A}^{2}\leq
V_{A}^{\max }}|p_{A}+iq_{A}\rangle \langle p_{A}+iq_{A}|dp_{A}dq_{A} \\
\tilde{D}^{B} &=&\frac{1}{\pi }\int\limits_{p_{B}^{2}+q_{B}^{2}\leq
V_{A}^{\max }}|p_{B}+iq_{B}\rangle \langle p_{B}+iq_{B}|dp_{B}dq_{B}
\end{eqnarray*}%
The filter $P_{A}^{d_{A}}$ and $P_{B}^{d_{B}}$ can be chosen in photon
number space. We let
\begin{eqnarray*}
P_{A}^{d_{A}} &=&|0\rangle _{A}\langle 0|+|1\rangle _{A}\langle
1|+...+|d_{A}-1\rangle _{A}\langle d_{A}-1| \\
P_{B}^{d_{B}} &=&|0\rangle _{B}\langle 0|+|1\rangle _{B}\langle
1|+...+|d_{B}-1\rangle _{B}\langle d_{B}-1|
\end{eqnarray*}%
where $|i\rangle _{A}\langle i|$ and $|j\rangle _{B}\langle j|$\ denote the
photon number state. Now we utilize theorem 1 to discuss the difference
between protocol 1 and protocol 2. We see that
\begin{eqnarray}
&&\sum_{i=0,j=d_{A}-1}^{\infty ,\infty }|_{A}\langle i|\tilde{D}%
^{A}|j\rangle _{A}|  \notag \\
&=&\frac{1}{\pi }\int\limits_{p_{A}^{2}+q_{A}^{2}\leq V_{A}^{\max
}}\sum_{i=0,j=d_{A}-1}^{\infty ,\infty }|_{A}\langle i|p_{A}+iq_{A}\rangle
\notag \\
&&\langle p_{A}+iq_{A}|j\rangle _{A}|dp_{A}dq_{A}  \notag \\
&=&2\int\limits_{r_{A}^{2}\leq V_{A}^{\max }}\sum_{i=0,j=d_{A}-1}^{\infty
,\infty }\frac{r_{A}^{i}r_{A}^{j}\exp [-r_{A}^{2}]}{\sqrt{i!j!}}dr_{A}
\notag \\
&=&2\int\limits_{r_{A}^{2}\leq V_{A}^{\max }}\sum_{i=0}^{\infty }\frac{%
r_{A}^{i}\exp [-r_{A}^{2}]}{\sqrt{i!}}\sum_{j=d_{A}-1}^{\infty }\frac{%
r_{A}^{j}}{\sqrt{j!}}dr_{A}  \label{overlap}
\end{eqnarray}%
where in the forth line we used the result that $|_{A}\langle
i|p_{A}+iq_{A}\rangle |=\frac{r_{A}^{i}\exp [-r_{A}^{2}/2]}{\sqrt{i!}}$ and
let $r_{A}^{2}=$ $p_{A}^{2}+q_{A}^{2}$. Under the case that $d_{A}\gg
V_{A}^{\max }$, we can use the Stirling formula to approximate $\sqrt{j!}$.
Then we have $\sum_{j=d_{A}}^{\infty }\frac{r_{A}^{j}}{\sqrt{j!}}\varpropto
(r_{A}/\sqrt{d_{A}})^{d_{A}}$, which exponentially goes to zero as $d_{A}$
increases. Then the whole term $\sum_{i=0,j=d_{A}}^{\infty ,\infty
}|_{A}\langle i|\tilde{D}^{A}|j\rangle _{A}|$ will exponentially go to zero
with the increase of $d_{A}$. By the same way we can prove that the term $%
\sum_{i=1,j=d_{B}}^{\infty ,\infty }|_{B}\langle i|\tilde{D}^{B}|j\rangle
_{B}|$ will also exponentially goes to zero with the increase of $d_{B}$.
Finally, for a given $\varepsilon ^{3}$ and large enough $N$, we can find a $%
d_{A}\ll N$ and $d_{B}\ll N$, that satisfy
\begin{equation*}
\sum_{i=1,j=d_{A}}^{\infty ,\infty }|_{A}\langle i|\tilde{D}^{A}|j\rangle
_{A}|+\sum_{i=1,j=d_{B}}^{\infty ,\infty }|_{B}\langle i|\tilde{D}%
^{B}|j\rangle _{B}|:=Err\leq \ \frac{\varepsilon ^{3}}{N}
\end{equation*}%
Then from the theorem 1 and 2 we know that the security of this CVQKD\
scheme can be approximated by the security of a scheme of dimension $%
(d_{A}d_{B})^{2}\ll N$ with errors no larger than $5\varepsilon $ ($Err$
exponentially approach to zero with the increase of $d_{A}$ and $d_{B}$, so
that $d_{A}$ and $d_{B}$\ are proportional with $\log (N/\varepsilon ^{3})$.
Then for large enough $N$, we can have $(d_{A}d_{B})^{2}\ll N$). Then $%
5\varepsilon +\epsilon $-secure secret key rate of heterodyne
detection CVQKD\ can be given by $2\varepsilon +\epsilon $-secure
secret key rate of protocol-2,
where Alice and Bob respectively put filters $P_{A}^{d_{A}}$ and $%
P_{B}^{d_{B}}$ before their detectors. As $N\rightarrow \infty $, we
can find large enough $(d_{A}d_{B})^{2}\ll N$, that allow
$\varepsilon \rightarrow 0$, and the security parameter $\epsilon $\
can goes to zero
From the Ref. \cite{Renner thesis} we know that, under the case that $%
(d_{A}d_{B})^{2}\ll N\rightarrow \infty $, the collective attack is optimal
for protocol 2 and its secret key rate can be given by that under collective
attack. Since the secrete key rate against collective attack of protocol 2
is no larger than that of protocol 1, under the infinite key size case the
unconditional secret key rate of heterodyne detection CVQKD\ equal to its
secret key rate under collective attacks and the collective is optimal.
Here, we require Alice and Bob give up such data whose amplitude is larger
than $V_{A}^{\max }$ and $V_{B}^{\max }$. We can expect that for large $%
V_{A}^{\max }$ and $V_{B}^{\max }$, the proportion of given up data is
extremely small. Such procedure only causes extremely small change of state,
and thus only cause extremely small change of security. On the other hand, a
realistic security proof for CVQKD\ should take such cut off procedure into
account. After all, in a realistic situation, the maximum value of
measurement results is always finite. \

\subsection{Unconditional security of DPSQKD}

Now we apply our result to coherent state DPSQKD, whose dimension is
infinite in principle. Up to now, the security
against collective attack for DPSQKD under noiseless case is proved \cite%
{DPSQKD}. Here we show that that proof actually is unconditional security
proof. To allow Alice and Bob do random permutation, in Ref. \cite{DPSQKD}
Zhao et al. cut the long sequence of coherent states into blocks and
regarded one block as one big state. Then Alice and Bob can permute these
big states. In the DPSQKD Alice sends Bob a big state $|\Psi _{\vec{x}%
}^{N_{b}}\rangle ={\bigotimes\limits_{i=1}^{N_{b}}}|(-1)^{x_{i}+1}\alpha
\rangle $ (denotes the state of a block), according to her binary string $%
\vec{x}=(x_{1},x_{2},...,x_{N_{b}})$, where $|(-1)^{x_{i}+1}\alpha
\rangle $ is a coherent state. Then Bob measures the phase
difference between each two individual state. The collective attack
means Eve attack these big states (blocks) independently with the
same method. Here we require Bob use the photon number resolving
detector. After many rounds of quantum communications, Bob announces
the maximum photon number received from one big state (one block).
Then if Bob put a filter that filters out all the state whose photon
number is larger than certain criteria, the measured results should
not change too much.

We see that if the efficiency of photon number resolving detector is
100\%, then we can definitely know the actual dimension of Bob's
received state. However, if that efficiency is not 100\%, we cannot
determine the exact dimension of Bob's state from the measured
photon numbers.

Here we discuss the imperfect detector case. In Ref. \cite{POVM2},
the POVM element of ineffective photon number resolving detector is
given. In that reference, the spacial mode of received photon state
has not been considered. If we take the spacial mode and other
components into account, we can extend that POVM element
corresponding to $n$ photons to be
\begin{equation}
\Pi _{n}=\sum\limits_{m=n}^{\infty }C^{n}_{m}\gamma^n(1-\gamma
)^{m-n}P_{m}\label{Pi}
\end{equation}
where $\gamma $ denotes detector efficiency and $P_{m}$ denotes the
projector to $m$ photon number subspace. We assume the dimension of
$m$ photon number subspace is $f_{m}$, and $P_{m}$ to be
\begin{equation}
P_{m}=\sum_{k=1}^{f_{m}}|\varphi_{k}^{m}\rangle\langle\varphi_{k}^{m}|
\label{pm}
\end{equation}
where $|\varphi_{k}^{m}\rangle$ denotes the orthogonal state of $m$
photon number subspace. It can be prove that $f_{m}\leq
l(m+l-1)!/m!$, where $l$ denotes the block size.

If Bob's maximum received photon number is $n_{0}$, then the POVM
element corresponding to this event can be given by
\begin{equation}
\tilde{D}^{B}=\sum_{n=0}^{n=n_{0}}\Pi_{n} \label{tildeD}
\end{equation}
In DPSQKD, if the block size is $l$, the dimension of Alice's
modulation is $2^{l}$, which is finite. Therefore we only need to
discuss Bob's state. We can construct Bob's filter to be
\begin{equation*}
P_{B}^{d_{B}}=\sum_{m=0}^{m=m_{0}}P_{m}
\end{equation*}
where $P_{m}$ is given by Eq. (\ref{pm}). Now we can use theorem 1
to estimate the difference between protocol 1 and protocol 2. We
enumerate the basis of the filter by $|\varphi_{k}^{m}\rangle$. Then
we have
\begin{eqnarray}
Diff:=\sum_{m=0,m'=m_{0},k,k'} \langle
\varphi_{k}^{m}|\tilde{D}^{B}|\varphi_{k'}^{m'}\rangle &&\\
\notag \leq \sum_{n=0}^{n_{0}}\gamma^n
\sum_{m=m_{0}}C_{m}^{n}(1-\gamma)^{m-n}l(m+l-1)!/m! \\ \notag \leq
\sum_{n=0}^{n_{0}}\gamma^n
\sum_{m=m_{0}}(1-\gamma)^{m-n}l/n!(m+l-1)^{l+n-1} \label{sum}
\end{eqnarray}
where in the second line we have used Eqs. (\ref{Pi}), (\ref{pm})
and (\ref{tildeD}) and the fact that $f_{m}\leq l(m+l-1)!/m!$ and in
the third line we used the fact that $m(m-1)...(m-n)\leq m^{n}$. It
can be seen that $Diff$ exponentially goes to zero as $m_{0}$
increases. Then for a given security parameter we can find a large
enough key size $N$ that gives the required security.

It also can be seen that if Bob use the perfect photon number
resolving detector or a detector that can given the upper bound of
the number of received photons (e.g. bourn up if received photon
number is too high), then they can find a protocol 2 that is exactly
same as protocol 1. Then we can immediately get a conclusion that
the collective attack is optimal under the infinite key size case.

\section{Conclusion:}

In the above we give a method to apply current exponential de
Finetti theorem to realistic QKD. In realistic QKD, the number of
Alice and Bob received photons is always finite and their
measurement results always belong to a finite region. This property
allow us effectively describe the QKD protocol in a finite
dimensional subspace with sufficiently small error. In this paper,
we introduce another finite dimensional protocol by putting finite
dimensional filters before the detectors, and shown the security
difference between the original unknown-dimensional protocol and
this finite dimensional protocol based on measurement results. Since
the security of that finite dimensional protocol is covered by
current information theoretical security proof method, the security
of a realistic unknown dimensional system can be solved. Our result
can be used to prove the unconditional security of heterodyne
detection CVQKD and DPSQKD. Finally, we prove that for heterodyne
detection CVQKD and DPSQKD collective attack is optimal under the
infinite key size case. The difference between protocol 1 and
protocol 2 will be meaningful if we consider the finite key size
case.

\textbf{Acknowledgement:} Special thanks are given to R. Renner for fruitful
discussions. This work is supported by National Natural Science Foundation
of China under Grants No. 60537020 and 60621064.

\appendix

\section{Detailed proof for Theorem 1}

At first we can see that $tr_{AB}[\tilde{D}^{\otimes N}\overline{%
(P_{AB}^{d_{A}d_{B}})^{\otimes N}}\rho _{A^{N}B^{N}}\overline{%
(P_{AB}^{d_{A}d_{B}})^{\otimes N}}]$ is no larger than $\max_{|\Psi
^{N}\rangle }\langle \Psi ^{N}|\tilde{D}^{\otimes N}|\Psi ^{N}\rangle $
where $|\Psi ^{N}\rangle \in \overline{(P_{AB}^{d_{A}d_{B}})^{\otimes N}}$.
To find $\max_{|\Psi ^{N}\rangle }\langle \Psi ^{N}|\tilde{D}^{\otimes
N}|\Psi ^{N}\rangle $, we need to expand the space $\overline{%
(P_{AB}^{d_{A}d_{B}})^{\otimes N}}$ by product spaces $\overline{%
P_{A}^{d_{A}}}(P_{A}^{d_{A}})^{\otimes N-1}(P_{B}^{d_{B}})^{\otimes N},...$.
Here, we let $P_{A_{k}}^{d_{A}}$ ($P_{B_{k}}^{d_{B}}$) denote the projector
to Alice's (Bob's) $k$-th state. Also we distinguish bases of $k$-th state
of Alice (Bob) as $|1\rangle _{A_{k}},|2\rangle _{A_{k}},...$, ($|1\rangle
_{B_{k}},|2\rangle _{B_{k}},...$). We know that $I_{A_{k}}=P_{A_{k}}^{d_{A}}+%
\overline{P_{A_{k}}^{d_{A}}}$ and $I_{B_{k}}=P_{B_{k}}^{d_{A}}+\overline{%
P_{B_{k}}^{d_{A}}}$, where $I_{A_{k}}$ and $I_{B_{k}}$ are the identity
matrixes corresponding to Alice and Bob's $k$-th states.\ Then we have
\begin{gather}
\langle \Psi ^{N}|\tilde{D}^{\otimes N}|\Psi ^{N}\rangle =\langle \Psi
^{N}|(P_{A_{1}}^{d_{A}}+\overline{P_{A_{1}}^{d_{A}}})\tilde{D}^{\otimes
N}|\Psi ^{N}\rangle  \notag \\
=\langle \Psi ^{N}|\overline{P_{A_{1}}^{d_{A}}}\tilde{D}^{\otimes N}|\Psi
^{N}\rangle +\langle \Psi ^{N}|P_{A_{1}}^{d_{A}}\tilde{D}^{\otimes N}|\Psi
^{N}\rangle  \notag \\
=C_{F}^{1}+C_{L}^{1}  \label{1}
\end{gather}%
where $C_{F}^{1}$ and $C_{L}^{1}$ respectively denote the first and second
term in the second line. Since
\begin{equation*}
\begin{array}{c}
\overline{P_{A_{1}}^{d_{A}}}=|d_{A}+1\rangle _{A_{1}}\langle
d_{A}+1|+|d_{A}+2\rangle _{A_{1}}\langle d_{A}+2|+... \\
I_{A_{1}}=|1\rangle _{A_{1}}\langle 1|+|2\rangle _{A_{1}}\langle 2|+...%
\end{array}%
\end{equation*}%
the $C_{F}^{1}$ can be given by%
\begin{eqnarray}
C_{F}^{1} &=&\langle \Psi ^{N}|\overline{P_{A_{1}}^{d_{A}}}\tilde{D}%
^{\otimes N}I_{A_{1}}|\Psi ^{N}\rangle  \label{C1F} \\
&=&\sum_{m_{1}=d_{A}+1,m_{1}^{\prime }=1}^{\infty ,\infty }\langle m_{1}|%
\tilde{D}^{A}|m_{1}^{\prime }\rangle \cdot  \notag \\
&&\langle \Psi ^{N}|m_{1}\rangle (\tilde{D}^{A})^{\otimes N-1}(\tilde{D}%
^{B})^{\otimes N}\langle m_{1}^{\prime }|\Psi ^{N}\rangle  \notag
\end{eqnarray}%
where we have used the fact that $P_{A_{k}}^{d_{A}}$ and
$\tilde{D}_{j}^{A}$ and $\tilde{D}_{j}^{B}$ are commutate if $k\neq
j$ and $\tilde{D}_{j}^{A}$ and $\tilde{D}_{j}^{B}$ denote POVM
elements corresponding to $j$-th state. We know
there exist two pure states $|\Phi _{1}^{m_{1}}\rangle $ and $|\tilde{\Phi}%
_{1}^{m_{1}^{\prime }}\rangle $ that can let $\langle m_{1}^{\prime }|\Psi
^{N}\rangle $ and $\langle m_{1}|\Psi ^{N}\rangle $\ be written as $\langle
m_{1}|\Psi ^{N}\rangle =\lambda _{1}|\Phi _{1}^{m_{1}}\rangle $ and $\langle
m_{1}^{\prime }|\Psi ^{N}\rangle =\lambda _{1}^{\prime }|\tilde{\Phi}%
_{1}^{m_{1}^{\prime }}\rangle $, where $|\lambda _{1}|\leq 1$ and $|\lambda
_{1}^{\prime }|\leq 1$. Then $C_{F}^{1}$ can be given by
\begin{eqnarray}
C_{F}^{1} &=&\langle \Psi ^{N}|\overline{P_{A_{1}}^{d_{A}}}\tilde{D}%
^{\otimes N}I_{A_{1}}|\Psi ^{N}\rangle  \label{C1F2} \\
&=&\lambda _{1}\lambda _{1}^{\prime }\sum_{m_{1}=d_{A}+1,m_{1}^{\prime
}=1}^{\infty ,\infty }\langle m_{1}|\tilde{D}^{A}|m_{1}^{\prime }\rangle
\cdot  \notag \\
&&\langle \Phi _{1}^{m_{1}}|(\tilde{D}^{A})^{\otimes N-1}(\tilde{D}%
^{B})^{\otimes N}|\tilde{\Phi}_{1}^{m_{1}^{\prime }}\rangle  \notag
\end{eqnarray}

Before giving the upper bound to $C_{F}^{1}$, we will discuss the upper
bound of $|\langle \Phi _{1}^{m_{1}}|(\tilde{D}^{A})^{\otimes N-1}(\tilde{D}%
^{B})^{\otimes N}|\tilde{\Phi}_{1}^{m_{1}^{\prime }}\rangle |$. It
is known that arbitrary POVM element $M$ can be written into a
diagonal form. We assume an arbitrary $M$ can be written as
\begin{equation*}
M=a_{1}|\varphi _{1}\rangle \langle \varphi _{1}|+a_{2}|\varphi _{2}\rangle
\langle \varphi _{2}|+...
\end{equation*}%
where $|\varphi _{1}\rangle ,|\varphi _{2}\rangle ,...$are orthogonal bases,
and $a_{1},a_{2},...$\ are positive real numbers and satisfy $a_{i}\leq 1$.
We let $|\psi \rangle $\ and $|\psi ^{\prime }\rangle $\ are two arbitrary
states. Now we consider the following value for $|\psi \rangle $\ and $|\psi
^{\prime }\rangle $.
\begin{equation*}
|\langle \psi |M|\psi ^{\prime }\rangle |=|a_{1}\langle \psi |\varphi
_{1}\rangle \langle \varphi _{1}|\psi ^{\prime }\rangle +a_{2}\langle \psi
|\varphi _{2}\rangle \langle \varphi _{2}|\psi ^{\prime }\rangle +...|
\end{equation*}%
From the fact that
\begin{eqnarray}
|\langle \psi |\varphi _{1}\rangle |^{2}+|\langle \psi |\varphi _{2}\rangle
|^{2}+... &\leq &1  \label{condition} \\
|\langle \psi ^{\prime }|\varphi _{1}\rangle |^{2}+|\langle \psi ^{\prime
}|\varphi _{2}\rangle |^{2}+... &\leq &1  \notag
\end{eqnarray}%
we know that for arbitrary states $|\psi \rangle $\ and $|\psi
^{\prime }\rangle $ and POVM element $M$, it is always satisfied
that
\begin{eqnarray}
|\langle \psi |M|\psi ^{\prime }\rangle | &\leq &\sqrt{|a_{1}\langle \psi
|\varphi _{1}\rangle |^{2}+|a_{2}\langle \psi |\varphi _{2}\rangle |^{2}+...}
\label{result} \\
&\leq &\sqrt{|\langle \psi |\varphi _{1}\rangle |^{2}+|\langle \psi |\varphi
_{2}\rangle |^{2}+...}\leq 1  \notag
\end{eqnarray}%
where in the first line we applied the Cauchy-Schwartz inequality which says
that
\begin{gather*}
|a_{1}b_{1}+a_{2}b_{2}+...|\leq \\
\sqrt{|a_{1}|^{2}+|a_{2}|^{2}+...}\sqrt{|b_{1}|^{2}+|b_{2}|^{2}+...}
\end{gather*}%
and in the second line we applied Eq. (\ref{condition}) and the fact that $%
a_{i}\leq 1$.\

Now we put Eq. (\ref{result}) into Eq. (\ref{C1F2}) and obtain
\begin{equation}
|C_{F}^{1}|\leq \sum_{m_{1}=d_{A}+1,m_{1}^{\prime }=1}^{\infty ,\infty
}|\langle m_{1}|\tilde{D}^{A}|m_{1}^{\prime }\rangle |  \label{in1}
\end{equation}

By the same way $C_{L}^{1}$ can be given by
\begin{eqnarray*}
C_{L}^{1} &=&\langle \Psi ^{N}|P_{A_{1}}^{d_{A}}\tilde{D}^{\otimes N}|\Psi
^{N}\rangle \\
&=&\langle \Psi ^{N}|P_{A_{1}}^{d_{A}}\overline{P_{A_{2}}^{d_{A}}}\tilde{D}%
^{\otimes N}|\Psi ^{N}\rangle +\langle \Psi
^{N}|P_{A_{1}}^{d_{A}}P_{A_{2}}^{d_{A}}\tilde{D}^{\otimes N}|\Psi ^{N}\rangle
\\
&=&C_{F}^{2}+C_{L}^{2}
\end{eqnarray*}%
where $C_{F}^{2}$ and $C_{L}^{2}$ respectively denote the first and the
second term in the second line.

As the $C_{F}^{1}$, the $C_{F}^{2}$ can be rewritten as%
\begin{eqnarray}
C_{F}^{2} &=&\langle \Psi ^{N}|P_{A_{1}}^{d_{A}}\overline{P_{A_{2}}^{d_{A}}}%
\tilde{D}^{\otimes N}I_{A_{2}}|\Psi ^{N}\rangle  \notag \\
&=&\sum_{m_{2}=d_{A}+1,m_{2}^{\prime }=1}^{\infty ,\infty }\langle m_{2}|%
\tilde{D}^{A}|m_{2}^{\prime }\rangle \cdot  \notag \\
&&\langle \Psi ^{N}|m_{2}\rangle P_{A_{1}}^{d_{A}}(\tilde{D}^{A})^{\otimes
N-1}(\tilde{D}^{B})^{\otimes N}\langle m_{2}^{\prime }|\Psi ^{N}\rangle
\label{CF2}
\end{eqnarray}%
Also there exist a pure state $|\Phi _{2}^{m_{2}}\rangle $ by which $\langle
\Psi ^{N}|m_{2}\rangle P_{A_{1}}^{d_{A}}$ can be written as $\lambda
_{2}\langle \Phi _{2}^{m_{2}}|$ and a pure state $|\tilde{\Phi}%
_{2}^{m_{2}^{\prime }}\rangle $ by which $\langle m_{2}^{\prime }|\Psi
^{N}\rangle $ can be given by $\lambda _{2}^{\prime }|\tilde{\Phi}%
_{2}^{m_{2}^{\prime }}\rangle $. Since $|\lambda _{1}|\leq 1$ and $|\lambda
_{2}|\leq 1$, from the Eqs. (\ref{result}) and (\ref{CF2})\ we know that
\begin{equation}
|C_{F}^{2}|\leq \sum_{m_{2}=d_{A}+1,m_{2}^{\prime }=1}^{\infty ,\infty
}|\langle m_{2}|\tilde{D}^{A}|m_{2}^{\prime }\rangle |  \label{in2}
\end{equation}%
If we continuously do such procedure, we will find that
\begin{equation}
\langle \Psi ^{N}|(\tilde{D}^{A}\tilde{D}^{B})^{\otimes N}|\Psi ^{N}\rangle
=\sum_{i=1}^{2N}C_{F}^{i}+C_{L}^{2N}  \label{f1}
\end{equation}%
and
\begin{equation}
|C_{F}^{i}|\leq \sum_{m_{i}=d_{A}+1,m_{i}^{\prime }=1}^{\infty ,\infty
}|\langle m_{i}|\tilde{D}^{A}|m_{i}^{\prime }\rangle |  \label{ini}
\end{equation}%
for $i\leq N$, and
\begin{equation}
|C_{F}^{i}|\leq \sum_{m_{i}=d_{B}+1,m_{i}^{\prime }=1}^{\infty ,\infty
}|\langle m_{i}|\tilde{D}^{B}|m_{i}^{\prime }\rangle |  \label{ini2}
\end{equation}%
for $i>N$, where
\begin{equation}
C_{L}^{2N}=\langle \Psi ^{N}|(P_{AB}^{d_{A}d_{B}})^{\otimes N}\tilde{D}%
^{\otimes N}|\Psi ^{N}\rangle =0  \label{f2}
\end{equation}%
and we have applied the fact that $|\Psi ^{N}\rangle \in $\ $\overline{%
(P_{AB}^{d_{A}d_{B}})^{\otimes N}}$. Finally from Eqs. (\ref{f1}), (\ref{ini}%
) (\ref{ini2}) and (\ref{f2})\ we can see that
\begin{gather}
|\langle \Psi ^{N}|\tilde{D}^{\otimes N}|\Psi ^{N}\rangle |\leq
\sum_{i=1}^{2N}|C_{F}^{i}|  \label{final} \\
\leq N[\sum_{i=1,j=d_{A}}^{\infty ,\infty }|_{A}\langle i|\tilde{D}%
^{A}|j\rangle _{A}|+\sum_{i=1,j=d_{B}}^{\infty ,\infty }|_{B}\langle i|%
\tilde{D}^{B}|j\rangle _{B}|]  \notag
\end{gather}%
Since Eq. (\ref{final}) holds for arbitrary $|\Psi ^{N}\rangle \in \overline{%
(P_{AB}^{d_{A}d_{B}})^{\otimes N}}$, the Theorem 1 is proved.


\begin{thebibliography}{99}
\bibitem{expl} Here we omit a small term proportional to $O[\text{log}(1/\epsilon
^{\prime })]$.
\bibitem{expl4} There may be many different purifications, but all
of them are different by local unitary transformations at Eve's
side. The local unitary transformation does not change the smooth
min-entropy. Therefore, all purifications are equivalent
actually.
\bibitem{expl2} Under the collective attack, the secret key rate can be
given by $\min_{\tilde{\rho}_{ABE}\in \Xi (\tilde{\rho}%
_{ABE})}H(X|E)-leak_{IR}$, where $\Xi (\tilde{\rho}_{ABE})$ is the
collection of all $\tilde{\rho}_{ABE}$s (in a finite dimensional
space) that consist with the observation. If we define a collection
$\Xi (\rho _{ABE})$\ to denote the collection of $\rho _{ABE}$s
(belong to a infinite dimensional
space) that consist with the observation, we can find that $\Xi (\tilde{\rho}%
_{ABE})\in \Xi (\rho _{ABE})$. Therefore we have $\min_{\tilde{\rho}%
_{ABE}\in \Xi (\tilde{\rho}_{ABE})}H(X|E)\geq \min_{\rho_{ABE}\in
\Xi (\rho _{ABE})}H(X|E)$.\ This inequality means that the secret
key rate against the collective attack of protocol 2 is no less than
that of protocol 1.
\end{thebibliography}
\end{document}